
%
\input phyzzx
\tolerance=1000
\voffset=-0.3cm
\hoffset=1.0cm
\sequentialequations
\def\rl{\rightline}

\def\r#1{$\bf#1$}

\def\t1{{\tilde 1}}

\def\SUSY{supersymmetry }

\def\NPB#1#2#3{Nucl. Phys. B {\bf#1} (19#2) #3}
\def\PLB#1#2#3{Phys. Lett. B {\bf#1} (19#2) #3}
\def\PRD#1#2#3{Phys. Rev. D {\bf#1} (19#2) #3}

\def\PRT#1#2#3{Phys. Rep. {\bf#1} (19#2) #3}

\def\l{\langle}
\def\r{\rangle}

\REF\UNI{U. Amaldi et al, \PRD{36}{87}{1385}; P. Langacker and M. Luo, \PRD{44}
{91}{817}; J. Ellis, S. Kelley and D. V. Nanopoulos, \PLB{249}{90}{441}.}
\REF\GSW{M. B. Green, J. H. Schwarz and E. Witten, {\it Superstring Theory,
Vols. 1 and 2} (Cambridge University Press, Cambridge 1987).}
\REF\KAP{V. Kaplunovsky, \NPB{307}{88}{145}.}
\REF\AD{K. R. Dienes and A. E. Faraggi, preprint IASSNS-HEP-95/24,
hep--th/9505018.}
\REF\JAP{T. Kobayashi, D. Suematsu and Y. Yamagishi, \PLB{329}{94}{27}.}
\REF\DIX{L. J. Dixon, V. Kaplunovsky and J. Louis, \NPB{355}{91}{649}.}
\REF\GCM{J. P. Deredinger, L. E. Ibanez and H. P. Nilles, \PLB{155}{85}{65};
M. Dine, R. Rohm, N. Seiberg and E. Witten, \PLB{156}{85}{55}.}
\REF\FFF{I. Antoniadis, C. Bachas, and C. Kounnas, \NPB{289}{87}{87};
I. Antoniadis and C. Bachas, \NPB{298}{88}{586};
H. Kawai, D.C. Lewellen, and S.H.-H. Tye, Nucl. Phys. B {\bf 288} (1987) 1.}
\REF\IBA{L. E. Ibanez, \PLB{318}{93}{73}.}
\REF\MOD{E. Halyo, \NPB{438}{95}{138}.}
\REF\THR{I. Antoniadis, J. Ellis, R. Lazace and D. V. Nanopoulos,
\PLB{268}{91}{188}.}
\REF\PDG{Particle Data Group, \PRD{50}{1994}.}
\REF\SUSY{A. Font, L. Ibanez, D. Lust and F. Quevedo, \PLB{245}{90}{401}.}
\REF\ANT{I. Antoniadis, \PLB{246}{90}{377}.}
\REF\RP{E. Halyo, Mod. Phys. Lett. {\bf A9} (1994) 1415.}
\REF\WNP{S. Ferrara, N. Magnoli, T. Taylor and G. Veneziano, \PLB{245}{90}
{409};D. L\"ust and T. Taylor, \PLB{253}{91}{335}.}
\REF\SQCD{D. Amati {\it et. al.}, \PRT{162}{88}{169}.}
\REF\EA{A. E. Faraggi and E. Halyo, preprint IASSNS-HEP-94/17, hep-ph/9405223.}
\REF\EDII{E. Halyo, preprint WIS-95/17/MAR-PH, hep--th/9505214.}
\REF\EDI{E. Halyo, \PLB{343}{95}{161}.}
\REF\KUM{D. Chang and A. Kumar, \PLB{211}{88}{76}; \PRD{38}{88}{1893}.}

\singlespace
\rl{WIS--95/36/AUG--PH}
\rl{\today}
\rl{T}
\pagenumber=0
\normalspace
\smallskip
\titlestyle{\bf{Gauge Coupling Unification due to Large String Threshold
Corrections}}
\smallskip
\author{Edi Halyo{\footnote*{e--mail address: jphalyo@weizmann.bitnet,
Address after September 1: Department of Physics, Stanford University,
Stanford CA 94305.}}}
\smallskip
\centerline {Department of Particle Physics}
\centerline {Weizmann Institute of Science}
\centerline {Rehovot 76100, Israel}
\vskip 2 cm
\titlestyle{\bf ABSTRACT}

We show that large string threshold corrections can reconcile the string and
MSSM unification scales in fermionic strings. This requires at least
three moduli with large VEVs which are different from each other and
MSSM states arising in an unconventional manner from the string spectrum. The
former is easily achieved by supersymmetry breaking by both hidden gaugino
and matter condensation whereas the latter needs to be seen in explicit
string models.

\singlespace
\vskip 0.5cm
\endpage
\normalspace

\centerline{\bf 1. Introduction}

The unification of the $SU(3) \times SU(2) \times U(1)$ gauge couplings is
considered to be a great achievement of the minimally supersymmetric
Standard Model (MSSM)[\UNI]. Assuming only the MSSM spectrum without any
additional states, one finds that the three gauge couplings unify around
$M_U \sim 2.5 \times 10^{16}~GeV$ with $g_U \sim 0.7$. This result
strenghtens the belief in both unification and supersymmetry.

The most fundamental theory of high energy physics to date is superstring
theory which gives a unified description of all interactions including
gravity[\GSW]. It is thus natural to try to realize the ideas of unification of
gauge couplings and supersymmetry in superstring theories. Whereas
supersymmetry
naturally arises from the superstring, the situation is different for
the unification of gauge couplings. In the string context, one assumes that
the three gauge couplings
unify around the string (or Planck scale) due to the fact that at that scale
all interactions are different aspects of the only string interaction.
In fact, it has been shown that gauge couplings of the string unify around
$M_S \sim g_S \times 5.5 \times 10^{17}~GeV$ where $g_S \sim 0.7$ is
the string coupling at that scale[\KAP]. Thus one finds that there is an order
of magnitude discrepancy between the MSSM and string unification scales which
needs to be explained. Conversely, if one takes the string unification scale
and runs down the gauge couplings with only the MSSM sprectrum, one obtains
results for $sin^2\theta(M_Z)$ and $\alpha_3(M_Z)$ which are in conflict with
experiments.

There have been different attempts to explain the above discrepancy.
Among them one can count introducing additional states at intermediate
energies[\AD], separating the
soft supersymmetry breaking masses around the TeV scale[\JAP] and including the
string threshold corrections[\KAP,\DIX]. The former two
depend on intermediate and low--energy physics (compared to the Planck scale)
and are therefore strongly model dependent. The string threshold
corrections, on the other hand, offer an elegant stringy solution to the
problem
around the Planck scale without invoking new particles and/or physics.
In addition, since these corrections are fixed for
a given string model they are much more constrained.
Unfortunately, in free fermionic (and orbifold) string models built
up to date, the threshold corrections are not large enough to solve the
problem due to the small (i.e. $\sim 1$) overall modulus
VEV which is a result of supersymmetry breaking by hidden gaugino condensation
[\GCM,\SUSY]. In addition, in free fermionic models, the sign of the
threshold corrections is wrong;
i.e. they increase the discrepance rather than decrease it[\AD].

In this letter, we consider unification of the gauge couplings due to
large string threshold corrections in fermionic strings[\FFF]. We show
that in order for these to reduce the string unification scale $M_S$ down
to $M_U$ or to give realistic values for $sin^2\theta_w(M_Z)$ and
$\alpha_3(M_Z)$, there must be at least three moduli with large ($>>1$) VEVs
which are different from each other. Moduli VEVs different than unity
correspond to a fermionic string deformed marginally by Abelian Thirring
interactions whose couplings are related to the VEVs themselves.
In addition, the MSSM states must arise from
the different sectors of the string spectrum in an unconventional manner so
that
the threshold corrections have the correct sign.
In section 2, we briefly review string threshold corrections and why they do
not offer a solution in realistic free fermionic string models built so far.
In section 3, we review supersymmetry breaking in the presence of hidden
gaugino and matter condensation and
how this scenario results in large moduli VEVs as required. In section 4,
we show how to get the desired threshold corrections which give the
correct $sin^2\theta_w(M_Z)$ and $\alpha_3(M_Z)$ (or reduce the string
unification scale $M_S$ down to $M_U$).
Section 5 contains our conclusions.

\bigskip
\centerline{\bf 2. String Threshold Corrections}

The one--loop renormalization group equations (RGE) for the gauge couplings
including the string threshold corrections are given by[\KAP]
$${16 \pi^2 \over {g_a^2(\mu)}}=k_a{16 \pi^2 \over {g^2_S}}
+b_a log{M_S^2 \over \mu^2}+\Delta_a \eqno(1)$$
where $a=1,2,3$ corresponding to $U(1)_Y,SU(2)_L, SU(3)_C$ respectively.
$b_a$ and $k_a$ are the MSSM $\beta$--function coefficients and the level of
the corresponding Kac--Moody algebras. In all string models
built to date, $k_2=k_3=1$ whereas $k_1$ can be considered as a free parameter
[\IBA]. $\Delta_a$ are the string
threshold corrections to the running of gauge couplings which arise from the
infinite tower of massive string states. They can be divided as
$\Delta_a= \tilde{\Delta}_a+c_a+Y$
where $\tilde{\Delta}_a$ gives the contributions which depend on the untwisted
moduli of the string model. The VEVs of these moduli are free in
perturbation theory to all orders but are fixed
by nonperturbative effects such as condensation in the hidden sector which
also break supersymmetry.

Untwisted moduli in free fermionic models have
been examined extensively in Ref. [\MOD]. Depending on the boundary condition
vectors defining the string model there can be up to three moduli of the $T$
and $U$ types, each pair corresponding to one of the the three compactified
tori. Every left--right asymmetric boundary condition or complex
world--sheet fermion eliminates two moduli so that there can be six, four, two
or no moduli in these models. In the section 4, we will see that
large threshold corrections which are phenomenologically acceptable require the
presence of at least one modulus from each torus. The string
spectrum can be divided into three different parts which have $N=4,2,1$
supersymmetry respectively. Only the sectors with $N=2$ supersymmetry
contribute to $\tilde{\Delta}_a$[\KAP,\DIX]. In realistic free fermionic
models there are three sectors with $N=2$ supersymmetry which
give one generation of fermions each[\AD].
These sectors also correspond to the twisted sectors of the string,
each giving one of the three twists of the $Z_2 \times Z_2$ which forms the
basis of realistic free fermionic strings.
The moduli dependent part of the threshold effects is given by
$$\tilde{\Delta}_a=-\sum_i {1\over 2} b_a^{\prime i}log[ReT_i|\eta(T_i)|^4
ReU_i|\eta(U_i)|^4] \eqno(2)$$
Here $b_a^{\prime i}$ are the $N=2$ $\beta$--function coefficients for the
gauge group fixed by $a$ and the $N=2$ sector $i=1,2,3$. $T_i$ in Eq. (2) are
the moduli corresponding to the torus left fixed under the twist which defines
the $N=2$ sector. $\eta(T)=e^{-\pi T/12} \Pi_n{(1-e^{-2\pi nT})}$ is the
Dedekind $\eta$--function. This one--loop
expression is exact due to the $N=2$ nonrenormalization
theorems. The small universal piece $Y$ turns out to be not important
since one can absorb it into the definition of $g_S$. The gauge dependent
piece $c_a$
does not depend on untwisted moduli and recieves contributions from
only the $N=2$ supersymmetric sector in free
fermionic strings. Generically $c_a$ are very small in free fermionic strings
[\THR] so that they do not affect the results for quantities such as
$sin^2\theta_w(M_Z)$ and $\alpha_3(M_Z)$ significantly. We neglect them in
the following. Using Eqs. (1) and (2) we find that
$$sin^2 \theta_w(M_z)={k_1 \over {k_1+k_2}}+{\alpha_1(M_Z) \over {4 \pi}}
{k_2 \over {k_1+k_2}}[A log ({M_Z^2 \over M_S^2})+\Delta_A] \eqno(3)$$
and
$$\alpha_3(M_Z)^{-1}={k_3 \over {k_1+k_2}}[{1\over \alpha_1(M_Z)}+{B\over {4
\pi}}log ({M_Z^2 \over M_S^2})+{\Delta_B \over {4 \pi}}] \eqno(4)$$
where
$$A=(b_1 {k_2 \over k_1}-b_2) \qquad B=(b_1+b_2-b_3 {(k_1+k_2) \over k_3})
\eqno(5)$$
and
$$\Delta_A=-(\Delta_1{k_2 \over k_1}-\Delta_2) \qquad \Delta_B=-(\Delta_1+
\Delta_2-\Delta_3{(k_1+k_2) \over k_3}) \eqno(6)$$
Here $\alpha_1(M_Z)=(127.9 \pm 0.1)$ is the electromagnetic structure constant
at the weak scale and $b_{1,2,3}=11,1,-3$ are the $\beta$--function
coefficients for the MSSM spectrum. $k_2=k_3=1$ whereas $k_1$ is a free
parameter. From Eq. (1) for the running coupling constants we can also obtain
the unification scale, $M_T$, in the presence of threshold corrections
$$M_T=M_S \prod_i[\sqrt{ReT_i}|\eta(T_i)|^2 \sqrt{ReU_i}|\eta(U_i)|^2]
^{b^{i \prime}_2-b^{i \prime}_3/b_3-b_2} \eqno(7)$$

Neglecting the threshold corrections and using $g_S \sim 0.7$ and
$M_S \sim g_S \times 5.5 \times 10^{17}~GeV$  we get
$sin^2 \theta_w(M_Z)=0.2187$ and $\alpha_3(M_Z)=0.195$ which do not agree
with the experimental values $sin^2 \theta_w(M_Z)=0.2319 \pm 0.0005$
and $\alpha_3(M_Z)=0.120 \pm 0.007$[\PDG].
The problem of string unification can be formulated in two equivalent
ways: without the threshold corrections
either $M_T$($=M_S$) is an order of magnitude larger than
$M_U$ or equivalently, with the values of $M_S$ and $g_S$ given above,
one finds too small (large) a value for $sin^2 \theta_w(M_Z)$
($\alpha_3(M_Z)$). The latter is simply the result of the extra running of
the gauge couplings from $M_S$ which is an order of magnitude larger than
$M_U$.

Can the string threshold corrections make up for the difference?
Considering Eq. (2) for $\tilde{\Delta}_a$, it is usually assumed that there is
only an overall modulus $T$ for simplicity even though most string models have
more than one untwisted moduli as we mentioned above. The VEV of $T$
in $\tilde{\Delta}_a$ is fixed by hidden sector condensation effects which
also break supersymmetry. In the scenario
with supersymmetry breaking by hidden gaugino condensation one obtains
$T \sim 1$[\SUSY]. Using the relation $\sum_i{1\over 2}{b_a^{\prime i}}=b_a$
which holds for free fermionic strings[\ANT],
one finds that $\tilde{\Delta}_a$ are too small to make up for the difference
between $M_S$ and $M_U$. In addition, in free fermionic string models,
$\tilde{\Delta}_a$ are such that the sign of the
corrections $\Delta_A$ and $\Delta_B$ is
wrong; i.e. they decrease $sin^2 \theta_w(M_Z)$ and increase $\alpha_3(M_Z)$
rather than the opposite.

Equivalently, one finds that threshold corrections
give through Eq. (7), $M_T>M_S$ i.e. they increase $M_T$ rather than decrease
it
to $M_U$.
This has also been established by explicit numarical calculations of
$\tilde{\Delta}_a$ in fermionic strings. The reason for this lies in the fact
that in fermionic strings all matter have modular weights
(which are related to the R charges of matter fields[\RP]) equal to $-1$ under
target space duality of the overall modulus $T$.
This leads to the relation $\sum_i{1\over 2}{b_a^{\prime i}}=b_a$
which together with $b_a$ for the MSSM spectrum
give the wrong sign. Thus, in order to explain the discrepancy between $M_S$
and
$M_U$ or obtain experimentally acceptable values of $sin^2 \theta_w(M_Z)$ and
$\alpha_3(M_Z)$ only by string threshold corrections we need a) large VEVs for
moduli so that the magnitude of $\tilde{\Delta}_a$ is large b) switch
the sign of $\tilde{\Delta}_{A,B}$.

\bigskip
\centerline{\bf 3. Supersymmetry breaking due to gaugino and matter
condensation}

We have seen that the moduli VEVs must be large ($>>1$)
in order to obtain large string threshold corrections.
In this section, we show that if supersymmetry is broken by hidden matter
condensation in addition to hidden gaugino condensation the vacuum is at
large $ReT_i$ as required. This is in contrast to the pure gaugino condensation
case with only an overall modulus in which the vacuum is given by
$ReT \sim 1.22$ [\SUSY] which is not large enough.

When the hidden gauge group ($SU(N)$) of a superstring (or supergravity)
becomes strongly
interacting gaugino condensates, $Y^3$, form. If there is also hidden matter
($M_i$), as it is the case in generic string models, then matter condensates
$\Pi_{ij}=\l M_i \bar M_j\r$ form in addition
to $Y^3$. The effective superpotential which describes the low--energy
effective theory after condensation is given by[\WNP]
$$W_{eff}={1 \over {32\pi^2}}Y^3 log\{exp(32\pi^2S)[c \eta(T)]^{6N-2M}
Y^{3N-3M} det \Pi \}-tr A \Pi, \eqno (8)$$
where $c$ is a constant and $A$ is the hidden matter mass
matrix which must be nonsingular in order to have a stable vacuum[\SQCD].
$S$ and $T$ are the dilaton and overall modulus respectively. For
simplicity here we consider only one modulus. The
generalization to more than one modulus is straightforward.

Taking the flat limit $M_P \to \infty$ one eliminates the strongly interacting
condensates $Y^3$ and $\Pi$ and obtains the effective superpotential in terms
of
$S$ and $T$
$$W_{np}(S,T)=\Omega(S) h(T) [det A]^{1/N}, \eqno(9)$$
where
$$\eqalignno{&\Omega(S)=-N exp(-32\pi^2S/N), &(10a) \cr
             &h(T)=(32\pi^2e)^{M/N-1}[c \eta(T)]^{2M/N-6}. &(10b)}$$
$det A$ is generically given by[\EA,\EDII,\EDI]
$$det A=k (ReS)^{-r} \phi_j^{s_j} \eta(T)^t \qquad r,s,t>0, \eqno(11)$$
where the $S$ dependence is obtained from the relation $g^2=1/4ReS$, $k$ is a
constant and $\phi_j$ are Standard Model scalar singlets whose VEVs give mass
to the hidden matter. The power of $\eta(T)$ is fixed by the requirement that
individual mass terms be modular invariant.
Using Eq. (9) for the superpotential and the Kahler potential
$$K=-log(2ReS)-3log(2ReT)-\sum_j(2ReT)^{n_j}\phi_j \phi_j^
{\dagger} \eqno(12)$$
we obtain the effective scalar potential[\EDII]
$$\eqalignno{V_{eff}&={e^{-\phi_j \phi_j^{\dagger}/2ReT} \over {16ReS (ReT)^3
|\eta(T)|^{8 \pi d^{\prime}}}}
|[det A]^{1/N}|^2 \{|2ReS \Omega_S-\Omega-{2 \Omega r \over N}|^2 \cr
&+|\Omega|^2 \left({4 d^{\prime2}(ReT)^3 \over (3 ReT- \phi_i
\phi_i^{\dagger})}
|G_2(T)-{3 \over {2ReT d^{\prime}}}+{\phi_j \phi_j^{\dagger} \over {4 (ReT)^2
d^{\prime}}}|^2-3\right)\}. &(13)} $$
where $d^{\prime}=(6N-2M-t)/4 \pi N$.
Here $G_2$ is defined through the derivative of $\eta(T)$ as
$\partial \eta(T)/\partial T=- \eta(T) G_2(T)/4 \pi$.
Comparing $V_{eff}$ above to that of the pure gaugino case
we see that the effect of hidden matter condensates and their mass terms is
simply to change the function $\hat G_2(T)=G_2-\pi/ReT$ in the pure gaugino
case
to $G_2(T)-{3/ 2ReT d^{\prime}}+{\phi_j \phi_j^{\dagger}/ 4 (ReT)^2
d^{\prime}}$
where $d^{\prime}$ is fixed by the hidden gauge group ($N$),
the matter content of the hidden sector ($M$) and the hidden
mass terms ($t$) in Eq. (11).

The potential above was studied in detail in Ref. [\EDII]. The results are as
follows. As $M$ and/or $t$ increase (which corresponds to more and/or lighter
hidden matter) so that $d^{\prime}$ decreases, $T_R$ at
the minimum increases from $1.22$ which is the value
obtained from pure gaugino condensation. For example $d^{\prime}=1/2 \pi$
and $d^{\prime}=3/10 \pi$ give minima at $T_R=3.75$ and $T_R=5.00$
respectively.
$ImT$ at the minimum on the other hand depends very weakly on $d^{\prime}$
and is an integer. Therefore one can get a large modulus VEV
if $d^{\prime}$ is small enough, i.e. if there is enough hidden matter
which is light enough.

It is well--known that free fermionic strings are formulated at $T=1$, the
fixed
point of target space duality. The large moduli VEVs mentioned above can only
result if there are untwisted moduli in the string spectrum. In that case,
these moduli can obtain VEVs different than unity due to nonperturbative
effects in the low--energy supergravity model. This is equivalent to
deforming the free fermionic string marginally by adding Abelian Thirring
interactions to the string action[\KUM]. The moduli VEVs are related
to the couplings
of the Abelian Thirring operators which deform the fermionic string marginally.
Thus a fermionic string with $T \not=1$
arising from hidden gaugino and matter condensation corresponds to
a marginally deformed free fermionic string.

For simplicity we considered only one modulus above. Our results can be easily
generalized to the more realistic case with more than one modulus of
either $T$ or $U$ type. In the next section we will see that for
acceptable threshold corrections one needs at least three moduli.
When there are a number of moduli, matter fields carry modular weights which
correspond to each one of them. (With three moduli, the modular weights of
matter fields are cyclic permutations of
($-1/2,-1/2,0$) rather than $-1$.) As a result, the parameters $M$ and $t$ in
Eqs. (10b) and (11) are generalized trivially to
$M_i$ and $t_i$ for each modulus $T_i$ or $U_i$. It is obvious
that as long as $M_i \not =M_j$, $t_i \not=t_j$ for $i \not=j$,
different moduli will obtain different VEVs from the minimization of $V_{eff}$.
In addition, if the corresponding $d^{\prime}_i$ are
small enough the moduli VEVs will be large as required for large
threshold corrections.

\bigskip
\centerline{\bf 4. Gauge coupling unification due to large string threshold
corrections}

In the previous section we saw how to get large moduli VEVs. In this section
we find what is required in order to get threshold corrections of the
correct sign and magnitude so that they reconcile the difference between
$M_S$ and $M_U$. By the correct sign of threshold corrections we mean
$\Delta_A>0$ and $\Delta_B<0$.

If there is only one modulus, one cannot obtain the correct
sign due to the relation $\sum_i{1\over 2}{b_a^{\prime i}}=b_a$
and the fact that $\sqrt{ReT}|\eta(T)|^2<1$
for all $T$. Having more than one modulus does not solve the
problem either in realistic models built so far.
In these models, the three sectors with $N=2$ supersymmetry give one
generation of fermions each so that $b_1^{i \prime}=20/3$,
$b_2^{i \prime}=0$ and $b_3^{i \prime}=-2$ for $i=1,2,3$.
(In these models, the two Higgs bosons arise from the Neveu--Schwarz
sector. Otherwise $b_2^{i \prime}$ is different but this does not affect the
above conclusion.)
It turns out that these values do not give threshold corrections
of the correct sign either even if the moduli have large VEVs which are
different from each other.

The problem is related to the values of $b_a^{i \prime}$ which follow from
the equal distribution of MSSM matter in the three $N=2$ sectors.
Clearly one can obtain
different values for $b_a^{i \prime}$ for different distributions of matter
into the $N=2$ sectors. It turns out that a necessary condition for the
required corrections is at least one sector with $b_3^{i \prime}>0$
which is not the case in realistic models with the equal division above.
Consider now a model in which the MSSM states arise from the three $N=2$
sectors in the following (unconventional) manner:
$\{Q_i,u_i,d_i\},\quad \{L_1,e_1\},\quad \{L_2,L_3,e_2,e_3\}$
where each curly bracket denotes a sector. Note that the two Higgs fields
are not included. Here we assume that they arise from the Neveu--Schwarz
sector of the string as is the case in realistic models. In any case, their
presence in any one of the sectors does not change our results qualitatively
since they do not affect $b_3^{i \prime}$. This distribution gives
$$\eqalignno{&b_1^{1 \prime}=34/3, \quad b_1^{2 \prime}=3,
\quad b_1^{3 \prime}=17/3,&(14a) \cr
&b_2^{1 \prime}=-5, \quad b_2^{2 \prime}=3, \quad b_2^{3 \prime}=2, &(14b) \cr
&b_3^{1 \prime}=6, \quad b_3^{2 \prime}=-6, \quad b_3^{3 \prime}=-6 &(14c)}$$

{}From the expression for the threshold corrections, Eq. (2), we see that at
best there can be three independent contributions to each $\tilde{\Delta}_a$;
one from each of the three $N=2$ supersymmetric sectors.
This requires the presence of at least one modulus (of either $T$ or $U$ type)
from each of the three sectors. Substituting Eq. (2) and the
experimental values of $sin^2\theta_w(M_Z)$ and $\alpha_3(M_Z)$
into Eqs. (3) and (4) we get two equations with four unknowns,
$k,\tilde{\Delta}_1,\tilde{\Delta}_2,\tilde{\Delta}_3$. $\tilde{\Delta}_a$
can be traded with
the real three unknowns of the problem which are $T_1,T_2,T_3$. (The
number of unknowns can increase up to six if $U$ type moduli are present since
they can obtain VEVs different than the $T$ moduli.)

Here we would like to stress three points. First not for every distribition
of MSSM states into sectors
there is a solution even though there are two equations and four unknowns.
For all distributions with $b_3^{i \prime}<0$, $i=1,2,3$
and for many with $b_3^{i \prime}>0$ for some $i$, we find that there is no
solution since they require $\sqrt{ReT}|\eta(T)|^2>1$ which is not possible.
Second, it is very difficult (if not impossible) to find solutions if
there are less than three moduli from three different sectors.
Third, given a distribution of states into sectors
not for every value of $k_1$ there is a solution. For example,
for the above distribution there is no solution if $k_1<1.33$ since this
requires $\sqrt{ReT}|\eta(T)|^2>1$.

{}From Eq. (3) and (4) we find that for $k_1=1.35$
$$\eqalignno{&-0.03=-\tilde{\Delta}_1+1.35 \tilde{\Delta}_2 &(15a) \cr
             &-2.37=\tilde{\Delta}_2-\tilde{\Delta}_3 &(15b)}$$
We can translate these into equations for the moduli $T_{1,2,3}$ by
defining (one can include $U_i$ in this definition if they exist)
$$log[ReT_{1,2,3}|\eta(T_{1,2,3})|^4]={x,y,z} \eqno(16)$$
and using the coefficients $b_a^{i \prime}$. Then Eq. (15) reads
$$\eqalignno{&-0.03=-0.54x+2.78y+2.77z &(17a) \cr
             &-2.37=0.50x-1.50y-2.00z &(17b)}$$
Every solution of the this set of equations gives a set of moduli VEVs
which reconciles the string and MSSM unification scales (for the assumed
matter distribution and $k_1$).
Thus, for $k_1=1.35$, we obtain a family of solutions which is fixed by
any of $x,y,z$, say $z$. (Note that $x,y,z \leq -1.06$ by definition. The
maximum is obtained at the fixed point $T_i=1$.)
For example, taking $z=-2.15$
gives $y=-1.10$ and $x=-16.64$ whereas $z=-1.20$ gives $y=-1.60$ and
$x=-14.34$.
We find that large moduli VEVs are needed to obtain these values.
For example, the first set of solutions corresponds to $T_1 \sim18.6,
T_2 \sim 1.3, T_3 \sim 3.1$ whereas the second set is given by
$T_1 \sim 16.4, T_2 \sim 2.3, T_3 \sim 1.6$.
(If there are also $U$ moduli present in some sectors, the corresponding VEVs
for $T_i$ are smaller than the above values depending on the VEVs of $U$
moduli.)
We saw in the previous section that these large moduli VEVs can be naturally
obtained in a supersymmetry breaking scenario due to both hidden gaugino and
matter condensation.

Above we found that there is no solution for $k_1<1.33$ whereas very close to
this value (i.e. $k_1=1.35$) there is a realistic solution. As $k_1$ increases,
solutions continue to exist but they require very large values of $T_i$.
For example for $k_1=1.5$ we need at least one $T_i \sim 100$ which is
impossible to obtain in the supersymmetry breaking scenario considered above
since this requires $d^{\prime} \sim 1/200$ which cannot be obtained for
realistic values of the parameters $N,M,t$.
Thus for the MSSM matter distribution we considered above
realistic solutions which resolve the difference between $M_S$ and $M_U$
require loosely $1.33<k_1<1.45$. The lower bound arises from the conditions
$x,y,z<0$ whereas the loose upper bound is due to the fact that $d^{\prime}$
cannot be smaller than a minimal value around $1/120$.

We condidered a specific distribution of MSSM states into the three
$N=2$ sectors for concreteness. There are other similar distributions which
give solutions with the same properties as long as there is one sector
with $b_3^{i \prime}>0$.
For example, any different distribution of leptons into the sectors gives
another solution. Generically, we need a matter distribution with
$b_3^{i \prime}>0$ for some $i$. This will give a one parameter family ($z$
in our case) of solutions for all $k_1$ between some minimal and maximal
values ($1.33$ and $\sim 1.45$ in our case).

\bigskip
\centerline{\bf 5. Conclusions}

In this letter, we discuss a way to reconcile the MSSM and string unification
scales or to obtain acceptable $sin^2\theta_w(M_Z)$ and $\alpha_3(M_Z)$ from
the string (with only the MSSM spectrum) by considering string
threshold corrections. This requires the presence of at least three
untwisted moduli from the three $N=2$ supersymmetric sectors of the string
spectrum. The moduli VEVs must be large and different from each other in order
to get large corrections. Large moduli VEVs can be naturally obtained in
a scenario with gaugino and matter condensation leading to supersymmetry
breaking. This corresponds to a fermionic string which is marginally
deformed by Abelian Thirring interactions whose couplings are related to the
moduli VEVs. In addition,
MSSM states must be distributed into the $N=2$ sectors in an unconventional
manner to obtain threshold corrections of the correct sign. Such a distribution
must satisfy $b_3^{\prime i}>0$ for some sector $i$. We gave a representative
example of such a distribution with the required moduli VEVs. We found a one
parameter (any of the moduli) family of solutions for every value of
$k_1$ between some
maximal and minimal values depending on the matter distribution.
There are other distributions with at least one $b_3^{\prime i}>0$
which give solutions with similar properties. Whether any of required
distributions of states can arise in realistic free fermionic string models
needs to be seen in explicit model building attempts.
We assumed throughout the paper that the contributions to the threshold
corrections which do not depend on the moduli are negligible, i.e.
$c_a,Y<< \tilde{\Delta}_a$ which holds for known models. If this is not the
case the required moduli VEVs
will be larger or smaller depending on the sign of these terms.

\bigskip
\centerline{\bf Acknowledgements}
This work was supported by the Department of Particle Physics and a Feinberg
Fellowship.

\vfill
\eject

\refout
\vfill
\eject

\end
\bye